\documentclass[aps,twocolumn,showpacs,superscriptaddress,groupedaddress]{revtex4}  
\pdfoutput=1 
\usepackage{graphicx}  
\usepackage{dcolumn}   
\usepackage{bm}        
\usepackage{amssymb}   
\usepackage{amsmath}   

\def\Rch{\check{R}}
\def\Re{\mathrm{Re}}
\def\Ktilde{\widetilde{K}}
\def\Rtilde{\widetilde{R}}
\def\Ctilde{\widetilde{C}}
\def\Wtilde{\widetilde{W}}
\def\Im{\mathrm{Im}}

\begin{document}



\title{A quantum diffusion law}

\author{Urbashi Satpathi}
\affiliation{Raman Research Institute, C. V. Raman Avenue, Sadashivanagar, Bangalore 560080, India.}

\author{Supurna Sinha}
\affiliation{Raman Research Institute, C. V. Raman Avenue, Sadashivanagar, Bangalore 560080, India.}

\author{Rafael D. Sorkin}
\affiliation{Raman Research Institute, C. V. Raman Avenue, Sadashivanagar, Bangalore 560080, India.}
\affiliation{Perimeter Institute for Theoretical Physics, 31 Caroline Street North, Waterloo, ON N2L 2Y5, Canada.}

\date{\today}

\begin{abstract}
We analyse diffusion at low temperature by bringing the fluctuation-dissipation theorem (FDT) to bear on a physically natural, viscous response-function $R(t)$. The resulting diffusion-law exhibits several distinct regimes of time and temperature, each with its own characteristic rate of spreading. As with earlier analyses, we find logarithmic spreading in the quantum regime, indicating that this behavior is robust.  A consistent $R(t)$ must satisfy the key physical requirements of {\it\/Wightman positivity\/} and {\it\/passivity\/},  and we prove that ours does so.  We also prove in general that these two conditions are equivalent when the FDT holds. Given current technology, our diffusion law can be tested in a laboratory with ultra cold atoms.


\end{abstract}

\pacs{05.30.-d,05.40.-a,05.40.Jc,32.80.Pj}
\maketitle

\section{Introduction}

A Brownian particle suspended in a liquid subject to thermal
fluctuations undergoes diffusion. What happens as we lower the
temperature and scale down the size of the particle until we reach a
regime where the diffusion is driven primarily by quantum zero point
fluctuations?

The question of diffusion in the presence of quantum zero point
fluctuations received a surge of interest in connection with
gravitational wave detection \cite{grav}.  Since such detectors need to
work at high levels of precision, the analysis of the Brownian motion of
the detector's components (such as mirrors) naturally comes into play
\cite{saulson,gaby,jaffino}.
%

In the present paper, we revisit this question starting
--- as before in our earlier paper \cite{rafsup} ---
from the fluctuation dissipation theorem.
In contrast to Ref. \cite{rafsup}, however,
we consider here
a response function whose behavior at very short times has been changed
from a step function to one which is more consistent physically, and
also closer to a form which is realizable in the laboratory.
At the time when \cite{rafsup} was written, the predicted logarithmic
diffusion was not experimentally accessible, but now that it is becoming
so, it seems important to analyze a response function which is not only
more realistic but also fully self-consistent.

The key physical requirements here are {\it\/Wightman positivity\/} of
the position correlation function and {\it\/passivity\/}, which is
essentially a version of the second law of thermodynamics.  We define
these conditions and discuss their interrelationships, showing in
particular that
they are equivalent
when the
fluctuation-dissipation theorem is in force.  Unlike the
response-function assumed in our earlier analysis, our present $R(t)$
satisfies both Wightman positivity and passivity.
This discussion, which is entirely new in relation to \cite{rafsup},
demonstrates as well that $R(t)$, in addition to being natural from an
experimental standpoint, is in principle realizable {\it\/exactly\/} as
a quantum gaussian process.

In the last two decades, light--matter
interaction has given a new impetus to such questions, and
one can now cool
dilute atomic gases down to temperatures of the order of $ 100 nK $,
where the transition to quantum degenerate regime can be
observed \cite{Anderson,chu,Ketterle,Optical,Magnetic,Ramansideband}.
As we point out below,
recent advances in
experimental technique
have progressed to the point that quantal
diffusion effects should now be observable.

The paper is organized as follows.
In Sec II we
obtain analytically,
the mean square displacement that results
from our newer response function, relegating most of the
computational details to the Appendix. In Sec III we describe certain
positivity conditions that consistent correlation functions and response
functions must satisfy, and we relate them to each other, showing that
our more realistic response function does satisfy them. In Sec IV we
review our main findings and discuss experimental possibilities.

\section{Diffusion law from Fluctuation Dissipation Theorem: quantum diffusion for a realistic response function}
Our starting point is the fluctuation dissipation theorem (FDT),
which in the frequency domain can be stated as follows \cite{rafsup, balescu}:
\begin{eqnarray}
  \mathrm{Im} \widetilde{R}(\nu)=\frac{1}{\hbar} \tanh(\pi\beta\hbar\nu)\widetilde{C}(\nu)\label{e1}
\end{eqnarray}
where, $ \beta=\frac{1}{k_{B}T} $
and $ (\widetilde{\cdot} )$
denotes the
{\it\/conjugate-linear\/}
Fourier transform defined by
\begin{eqnarray*}
   \widetilde{f}(\nu) = \int dt \, e^{2\pi i\nu t} \, f^{*}(t)  \ .
\end{eqnarray*}
$ \widetilde{R}(\nu) $ and $ \widetilde{C}(\nu) $ are
respectively the
transforms of the time-dependent response-function ${R}(t)$ and of the
auto-correlation function ${C}(t)$:
\begin{eqnarray}
   R(t) &=& \frac{1}{i\hbar}\left\langle\left[ x(0),x(t)\right] \right\rangle \theta(t)\label{e11}\\
   C(t) &=& \frac{1}{2} \left\langle\left\lbrace x(0),x(t) \right\rbrace \right\rangle \label{e12}
\end{eqnarray}
where, $ x(t) $ is the displacement and $ \theta(t) $ is the unit step function defined as:
\begin{eqnarray*}
\theta(t)=
\begin{cases}
    1,&  t\geq 0\\
        0,& t<0
\end{cases}
\end{eqnarray*}

We now consider the fluctuation dissipation theorem in the time
domain. To that end, we first consider, instead of $ R(t) $, the
equivalent odd function \cite{rafsup},
\begin{eqnarray*}
  \Rch (t)= \mathrm{sgn}(t)R(\vert t\vert)\label{e13}
\end{eqnarray*}
where $ \mathrm{sgn}(t) $ is the sign or signum function, defined as:
\begin{eqnarray*}
\mathrm{sgn}(t)=
\begin{cases}
    1,&  t>0\\
        0,& t=0\\
        -1,& t<0
\end{cases}
\end{eqnarray*}
$ R(t) $ defined in Eq. (\ref{e11}) is a causal function which vanishes
for $ t<0 $, whereas $ \Rch (t) $ exists for the entire time domain.
This
 enables us to recast the fluctuation dissipation theorem as follows \cite{rafsup}:
\begin{eqnarray}
  \widetilde{\Rch }(\nu)= \frac{2i}{\hbar}\tanh(\pi \beta \hbar \nu)\widetilde{C}(\nu)
\end{eqnarray}
We finally arrive at \cite{rafsup},
\begin{eqnarray}
\hspace*{-0.3cm}
 C(t)=\frac{1}{2\beta}\int_{-\infty}^{\infty}dt' \mathrm{sgn}(t'-t)R(\vert t'-t \vert)\coth \left( \frac{\pi t'}{\beta\hbar}\right) + c \label{e2}
\end{eqnarray}
where $ c $ is a constant.

Our analysis focuses on the mean square displacement, and deduces it
from the position auto-correlation function.
The mean square displacement is given by,
\begin{eqnarray}
  \langle \Delta x^2 \rangle
  &=& \langle[x(t)-x(0)]^{2}\rangle = 2 \left[ C(0)-C(t)\right] \label{eq0}
\end{eqnarray}
Using Eq. (\ref{e2}), we can write \cite{rafsup},
\begin{eqnarray}
  \langle \Delta x^2 \rangle &=&
  \frac{1}{\beta}
  \int_{0}^{\infty} dt'R(t') \left\lbrace 2 \coth \left(\frac{t'}{t_{th}}\right) - \coth \left(\frac{t'+t}{t_{th}}\right) \right.  \nonumber\\
   &-& \left. \coth \left( \frac{t'-t}{t_{th}}\right) \right\rbrace \label{eq1}
\end{eqnarray}
where,
$t_{th}={\beta\hbar}/{\pi}$ is the thermal time.
%
The definition of mean square displacement in Eq. (\ref{eq0}) entails
that it is necessarily positive.
This condition further restricts the choice of the response function, as
we discuss more fully below.

We will consider primarily the following response-function:
\begin{eqnarray}
  R(t)&=&\mu \left( 1-e^{-\frac{t}{\tau}}\right)\theta\left( t\right)
  \label{eq2}
\end{eqnarray}
Here $\mu$ can be called the mobility and $\tau$ the relaxation time.
This response-function is suggested by the venerable model of a viscous
medium.  (Such a medium can be realized experimentally as a
three-dimensional ``optical molasses'' of the type used for laser
cooling of dilute atomic gases \cite{chu}.)
Although (\ref{eq2}) is not as easy to analyze as our earlier response
function, it has the important advantage of being fully self-consistent
physically, in the sense that it complies
with certain positivity criteria which we
discuss in detail in Sec III.

With this response function, Eq. (\ref{eq1}) reduces to (see Appendix
for details),
\begin{eqnarray}
\hspace*{-0.3cm}
 \langle \Delta x^2 \rangle
  &=&
 \frac{2\mu}{\beta } t_{th} \left\lbrace \ln\left[2\sinh\left(\frac{t}{t_{th}}\right)\right]
  + \psi^{0}\left(1 + \frac{t_{th}}{2\tau}\right)
  + \gamma + \right.\nonumber\\
  && \left.
  \frac{2\tau}{t_{th}}
  \left[{}_{2}F_{1}\left(1, \frac{t_{th}}{2\tau}, 1+\frac{t_{th}}{2\tau}, e^{-\frac{2 t}{t_{th}}}\right) - 1\right]
  \right\rbrace
  \label{eq4}
\end{eqnarray}
Here, $ \gamma\approx 0.5772$ is the Euler-Mascheroni constant,
$ \psi^{0} $ is a Polygamma function of order zero,
and $ {}_{2}F_{1} $ is a Hypergeometric Function (see Appendix).

\begin{figure}
\includegraphics[scale=0.36]{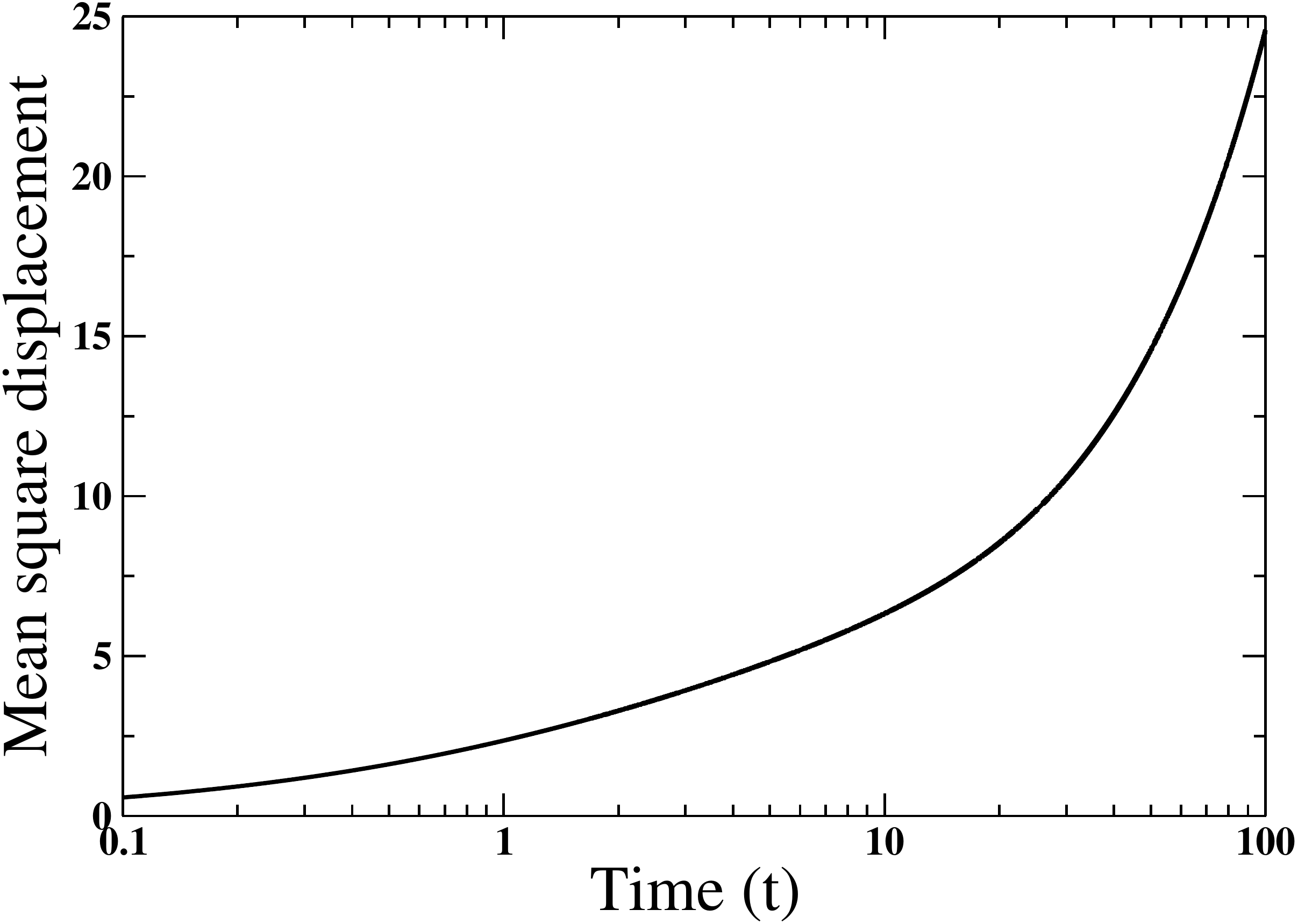}
\caption{\label{fig1}
 Plot of the mean square displacement as a function of time (in
 logarithmic scale) in arbitrary units, obtained from
 Eq.~(\ref{eq4}). In this case, the relaxation time is taken to be
 $ \tau=1 $ and the thermal time is taken to be $ t_{th}=10 $.}
\end{figure}
In Eq.~(\ref{eq4}), the Polygamma function and the Hypergeometric function are
always positive, but as $ t\rightarrow 0 $, the logarithm goes negative. This,
however, is counter-balanced by the Hypergeometric function, resulting in a net
positive value of the mean square displacement. We have checked this
semi-analytically and found that the R.H.S of Eq.~(\ref{eq4}) is always
positive. Our newer response function thus passes this consistency test.
(See Fig.~\ref{fig1} where we have plotted $ \langle \Delta x^2 \rangle$
against time over a large range of time scales using Eq.~(\ref{eq4}).)

We can identify several different
limiting cases or ``regimes'', depending on the three time scales:
$ \tau $ = relaxation time,
$ t_{th}$ = thermal time,
and $t$ = observation time.
The thermal time $ t_{th} $ is related to the
temperature $ T $ by $ \beta\hbar=\frac{\hbar}{k_{B}T}$.  Depending on
these time scales there can be six distinct possibilities, which we will
now discuss.

In Ref. \cite{rafsup}, the cases we will call 1, 2, and 3 were studied
and analytical forms for the mean square displacement were discussed for
a cruder step-function form of response function, $
R(t)=\mu\theta(t-\tau)$. Using the more realistic response function of
Eq.~(\ref{eq2}), it is possible to get analytical forms for the other
three cases as well (cases A,B, and C).


\subsection*{Case A: $t<<\tau<<t_{th}$}
\begin{figure}
\includegraphics[scale=0.36]{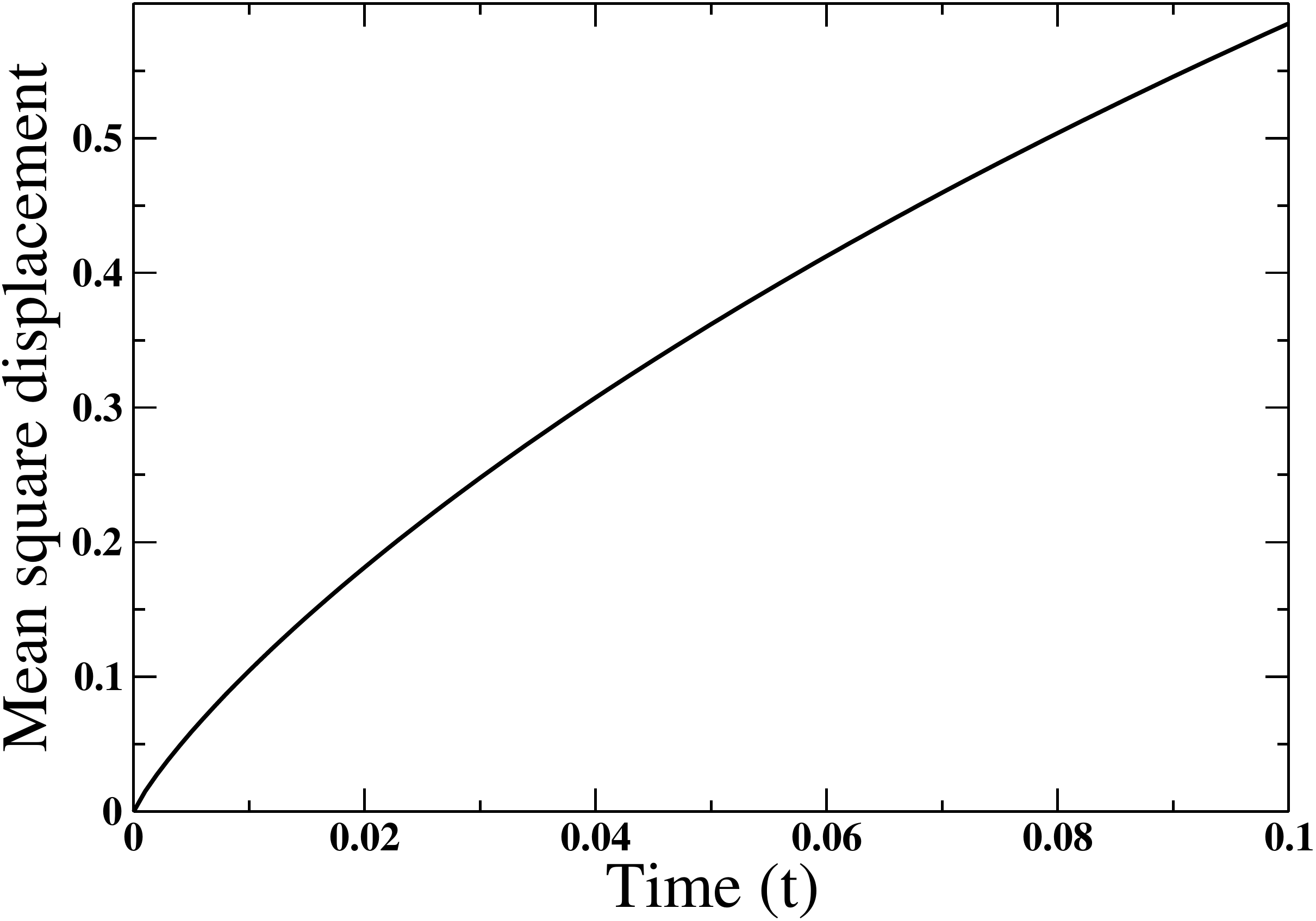}
\caption{\label{fig2}
 Plot of the mean square displacement as a function of time in arbitrary
 units, under the condition, $t<<\tau<<t_{th}$. The mean square
 displacement is obtained from Eq.~(\ref{q1}). In this case, the
 relaxation time is taken to be $ \tau=1 $ and the thermal time is taken
 to be $ t_{th}=10 $.}
\end{figure}
In this limit, using Eq.~(\ref{a13}) retaining terms to first order in $
t $, and Eq.~(\ref{a15}),
Eq.~(\ref{eq4}) reduces to,
\begin{eqnarray}
  \langle \Delta x^2 \rangle
   &=&
  \frac{2\mu}{\beta }
  \left\lbrace \frac{t_{th}t}{\tau}\left[1-\ln\left( \frac{t}{\tau}\right) -\gamma +\frac{\tau}{t_{th}} \right] \right\rbrace
  \label{q1}
\end{eqnarray}
In Fig. \ref{fig2}, we have plotted the mean square
displacement as a function of time using  Eq. (\ref{q1}).
It is possible to estimate the order of magnitude for the time and the
temperature in this regime. Considering the relaxation time for sodium
\cite{chu} to be $ \tau=10 \mu s $, $ t $ turns out to be a few $ ns $
and $ T $ is of the order of $ \mu K $.

\subsection*{Case B: $t<<t_{th}<<\tau$}
\begin{figure}
\includegraphics[scale=0.36]{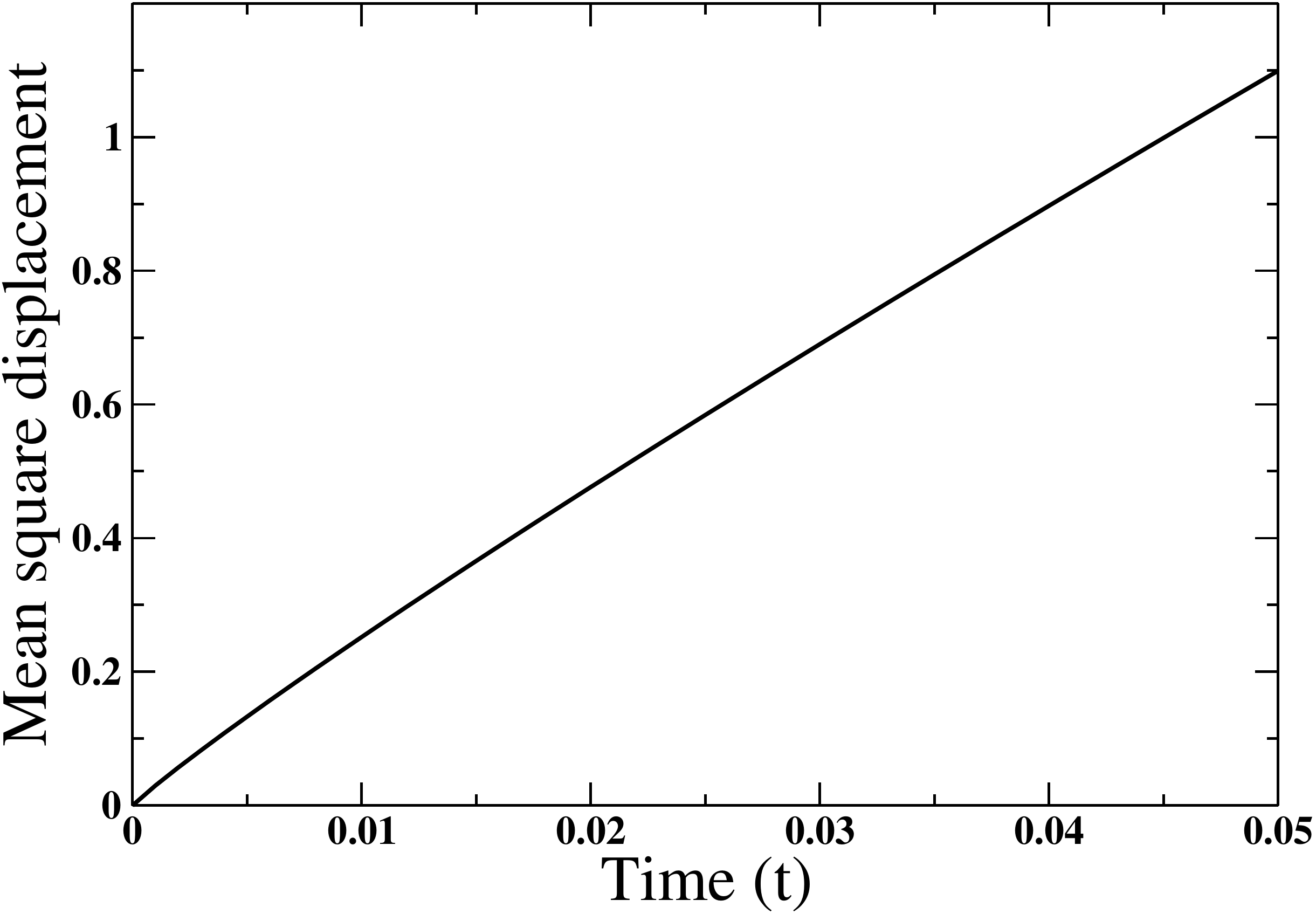}
\caption{\label{fig3}
 Plot of the mean square displacement as a function of time in arbitrary units, under the
 condition, $t<<t_{th}<<\tau$. The mean square displacement is obtained
 from Eq.~(\ref{q2}). In this case, the relaxation time is taken to be
 $ \tau=1 $ and the thermal time is taken to be $ t_{th}=0.1 $.}
\end{figure}
In this limit,
using Eq.~(\ref{a13}) retaining terms to first order in $ t $, and Eq.~(\ref{a16}),
Eq. (\ref{eq4}) reduces to,
\begin{eqnarray}
\hspace*{-0.6cm}
 \langle \Delta x^2 \rangle
  &=&
 \frac{2\mu}{\beta}\left\lbrace \frac{t_{th}t}{\tau}
 \left[1-\ln\left( \frac{2t}{t_{th}}\right) -\frac{\pi^2}{12}\frac{t_{th}}{\tau} +\frac{\tau}{t_{th}} \right]
 \right\rbrace
 \label{q2}
\end{eqnarray}

In Fig. \ref{fig3}, we have plotted the mean square
displacement as a function of time using  Eq. (\ref{q2}).
In this case, considering the same relaxation time, i.e. $ \tau=10 \mu s$, the observation time can be estimated to be of the order of $ ns $
and the temperature of the order of a few $ \mu K $ to $ mK $.

\subsection*{Case C: $t_{th}<<t<<\tau$}
\begin{figure}
\includegraphics[scale=0.36]{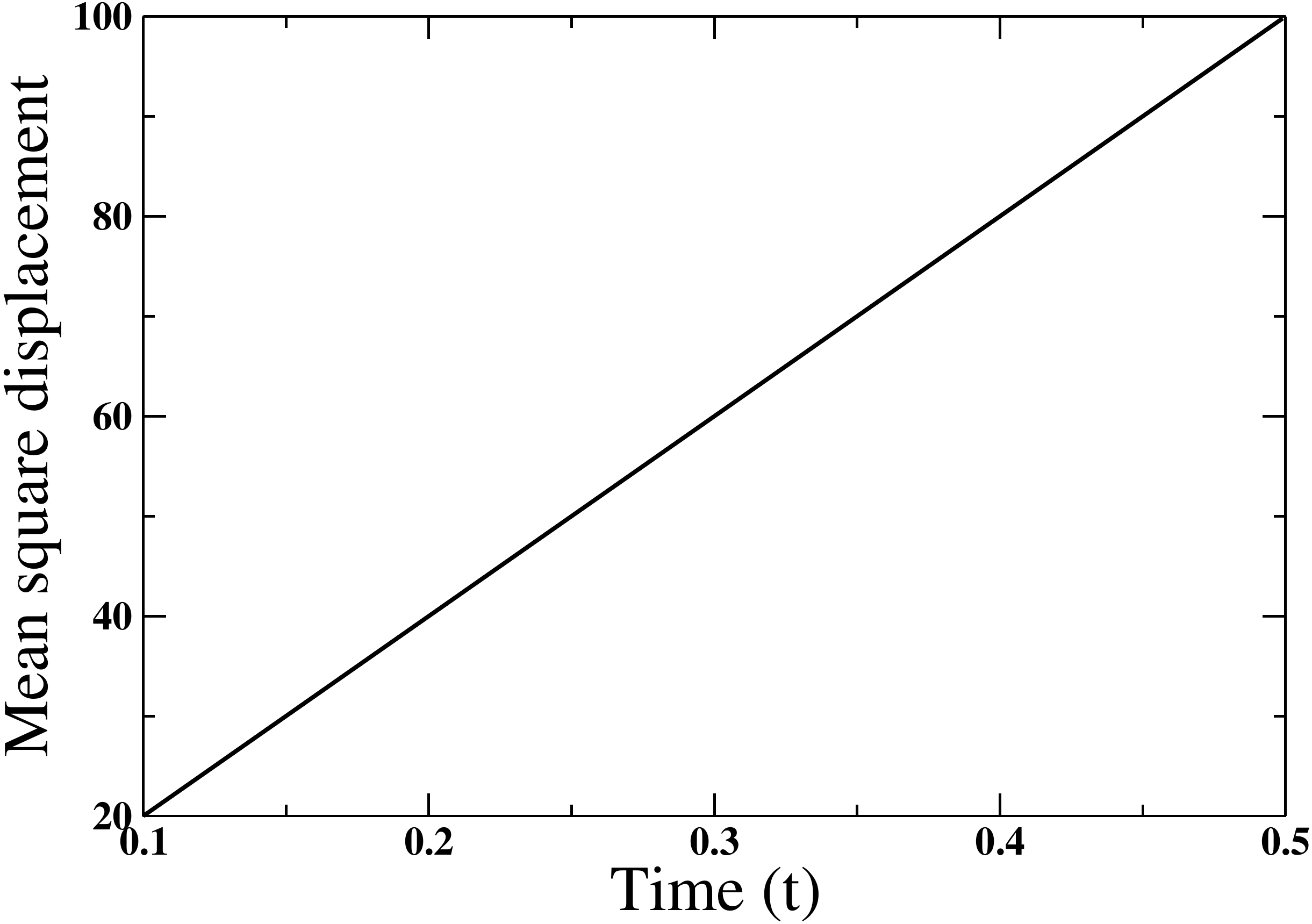}
\caption{\label{fig4}
 Plot of the mean square displacement as a function of time in arbitrary
 units, under the condition, $t_{th}<<t<<\tau$.  The mean square
 displacement is obtained from Eq.~(\ref{q3}).  In this case, the
 relaxation time is taken to be $ \tau=1 $ and the thermal time is taken
 to be $ t_{th}=0.01 $.}
\end{figure}
In this limit using Eq.~(\ref{a10a}), Eq.~(\ref{a10b}) and Eq.~(\ref{a16}),
Eq. (\ref{eq4}) reduces to,
\begin{eqnarray}
 \langle \Delta x^2 \rangle
  &=&\frac{2\mu}{\beta}
 \left\lbrace t + \frac{\pi^2}{12}\frac{t_{th}^2}{\tau} \right\rbrace
 \label{q3}
\end{eqnarray}
In Fig. \ref{fig4}, we have plotted the mean square
displacement as a function of time using Eq. (\ref{q3}).
In this case, considering the same relaxation time, i.e. $ \tau=10 \mu s$,
the observation time can be estimated to be of the order of a few $
ns $ to a few $ \mu s $ and the temperature of the order of a few $ mK $ to $ 1 K $.

\subsection*{Case 1: Quantum regime}
In the quantum limit, i.e., $\tau<<t<<t_{th}$,
using Eq.~(\ref{a10c}) and Eq.~(\ref{a16}),
Eq. (\ref{eq4}) reduces to,
\begin{eqnarray}
 \langle \Delta x^2 \rangle
  &=&
 \frac{2\mu}{\beta }t_{th}\left\lbrace \ln\left(\frac{t}{\tau}\right) + \gamma\right\rbrace
 \label{eq6}
\end{eqnarray}
For comparison,
with the step function response function,
the mean square displacement in the quantum domain was \cite{rafsup},
\begin{eqnarray}
  \langle \Delta x^2 \rangle &=&\frac{2\mu}{\beta } t_{th}\ln\left(\frac{t}{\tau}\right)
 \label{eq61}
\end{eqnarray}
\begin{figure}
  \includegraphics[scale=0.36]{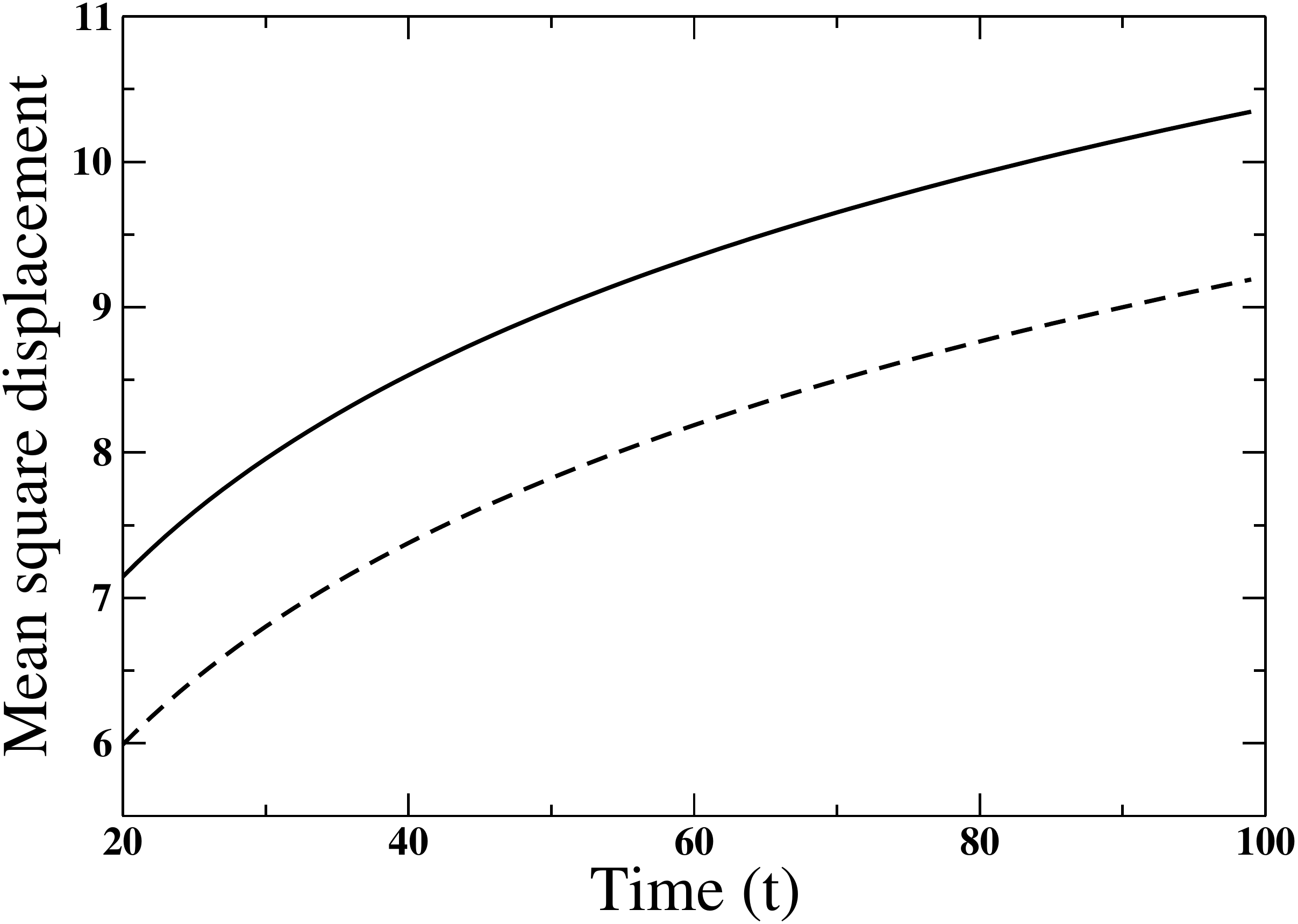}
 \caption{\label{fig5}
  Plots of the mean square displacement as a function of time under the condition, $\tau<<t<<t_{th}$.
  The solid line is the mean square displacement using Eq.~(\ref{eq6}),
  and the dashed line is the mean square displacement using Eq.~(\ref{eq61}).
  In this case, the relaxation time is taken to be $ \tau=1 $ and the thermal time is taken to be $ t_{th}=100 $.}
\end{figure}

In Fig. \ref{fig5}, we have plotted the mean square
displacement as a function of time, in the quantum domain. The two
curves are obtained using Eq. (\ref{eq6}) for the newer response function
and Eq. (\ref{eq61}) for the step-function response function. The two
curves qualitatively show the same logarithmic behaviour. But we notice
a quantitative difference as manifested in a difference in the size of
the intercept.

In this case, considering the same relaxation time, i.e. $ \tau=10 \mu s
$, the observation time can be estimated to be of the order of a few $
ms $ and temperatures of a few $ nK $ or below. Reaching this
temperature regime seems possible with present experimental techniques in
cold atom experiments where temperatures down to $ 500 pK $ can be
reached \cite{Ketterle}.

\subsection*{Case 2: Intermediate regime}
In the intermediate time regime, i.e., $\tau<<t_{th}<<t$,
using Eq.~(\ref{a10c}), Eq.~(\ref{a10d}) and Eq.~(\ref{a16}),
Eq. (\ref{eq4}) reduces to,
\begin{eqnarray}
\langle \Delta x^2 \rangle &=&\frac{2\mu }{\beta}\left\lbrace  t+t_{th}\left[\ln\left(\frac{t_{th}}{2\tau}\right)+\gamma\right]\right\rbrace
\label{eq7}
\end{eqnarray}
Using the step-function response function, the mean square displacement
in the intermediate time domain was \cite{rafsup},
\begin{eqnarray}
\langle \Delta x^2 \rangle &=&\frac{2\mu }{\beta}\left\lbrace  t+t_{th}\ln\left(\frac{t_{th}}{2\tau}\right)\right\rbrace\label{eq71}
\end{eqnarray}
\begin{figure}
  \includegraphics[scale=0.36]{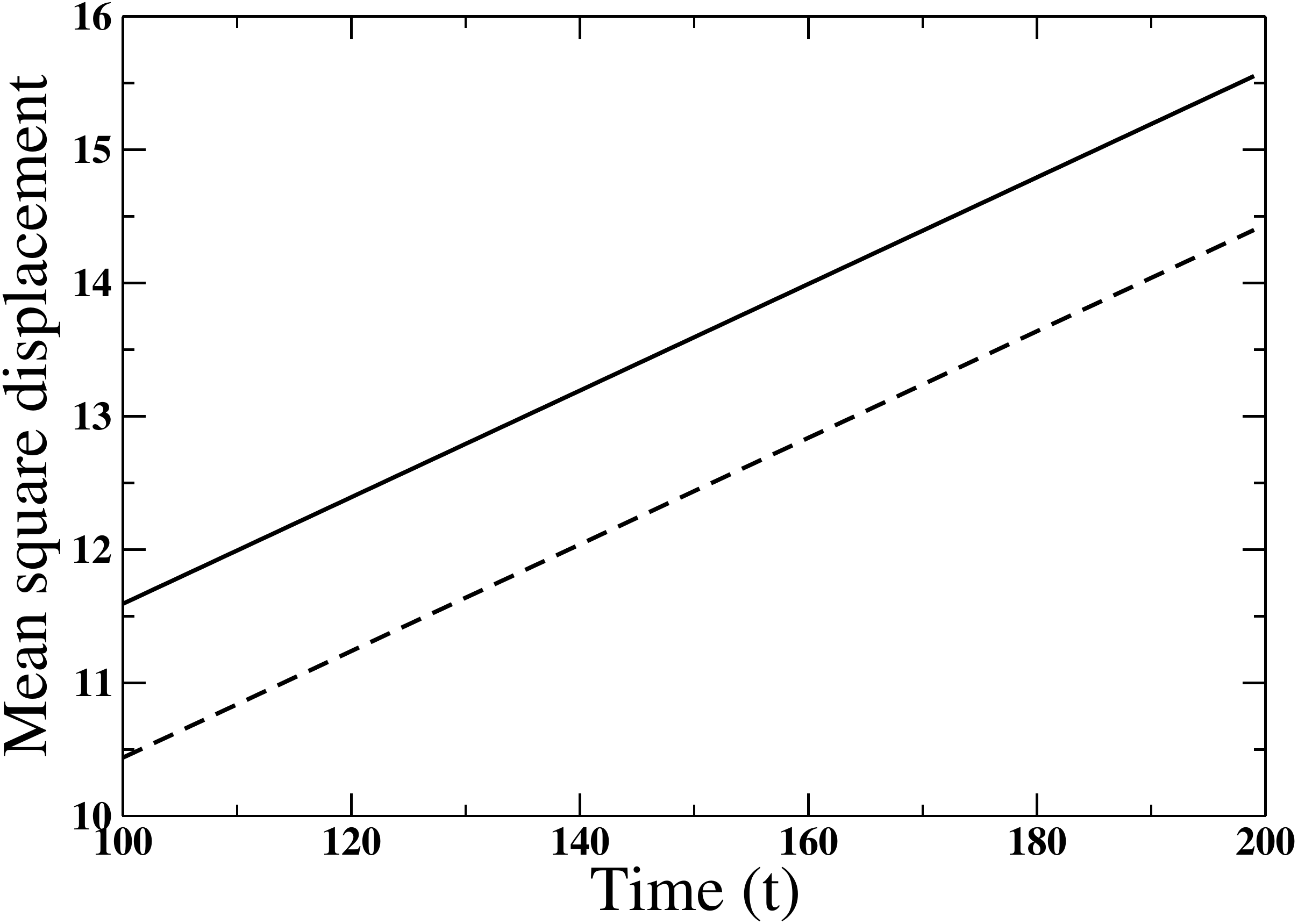}
  \caption{\label{fig6}
  Plot of the mean square displacement as a function of time in arbitrary units, under the
  condition, $\tau<<t_{th}<<t$. As in Fig. \ref{fig5}, the solid line
  is the mean square displacement using Eq.~(\ref{eq7}) and
  the dashed line is the mean square displacement using Eq.~(\ref{eq71}).
  In this case, the relaxation time is taken to be $ \tau=1 $ and the thermal time is taken to be $ t_{th}=50 $.}
\end{figure}
In Fig. \ref{fig6}, we have shown the plot of the mean square
displacement in the intermediate time regime. The two curves are
obtained using Eq. (\ref{eq7}) for the newer response function
and Eq. (\ref{eq71}) using step-function response function.
As in Case 1 we notice that the two curves show the same qualitative behaviour. There is, however, a
quantitative difference which is captured by the size of the intercept, as we noticed in Case 1.

In this case, considering the same relaxation time, i.e. $ \tau=10 \mu s$,
the observation time can be estimated to be of the order of a few $ m s $
and the temperature of the order of a few $ nK $ to $ \mu K $. This
regime can be easily realized with ultra cold atoms where the typical
relaxation times can be around a few $ \mu s $ and temperature regime of
$ 10 \mu K $ can be reached using laser cooling and a few tens of $ nK $
can be reached using evaporative cooling in optical \cite{Optical} or
magnetic traps \cite{Magnetic} or Raman side-band cooling in optical
lattices \cite{Ramansideband}.

\subsection*{Case 3: Classical regime}
In the classical limit, i.e., $t_{th}<<\tau<<t$,
using Eq.~(\ref{a10a}) and Eq.~(\ref{a10b}),
Eq. (\ref{eq4}) reduces to,
\begin{eqnarray}
 \langle \Delta x^2 \rangle
  &=&
 \frac{2\mu}{\beta}t_{th}\left\lbrace \ln\left[e^{\frac{t}{t_{th}}}\right] \right\rbrace
 =\frac{2\mu }{\beta}t \label{eq5}
\end{eqnarray}
Using the step-function response-function, the mean square displacement
in the classical domain was \cite{rafsup} the same as in Eq. (\ref{eq5}).
\begin{figure}
  \hspace*{-0.3cm}\includegraphics[scale=0.36]{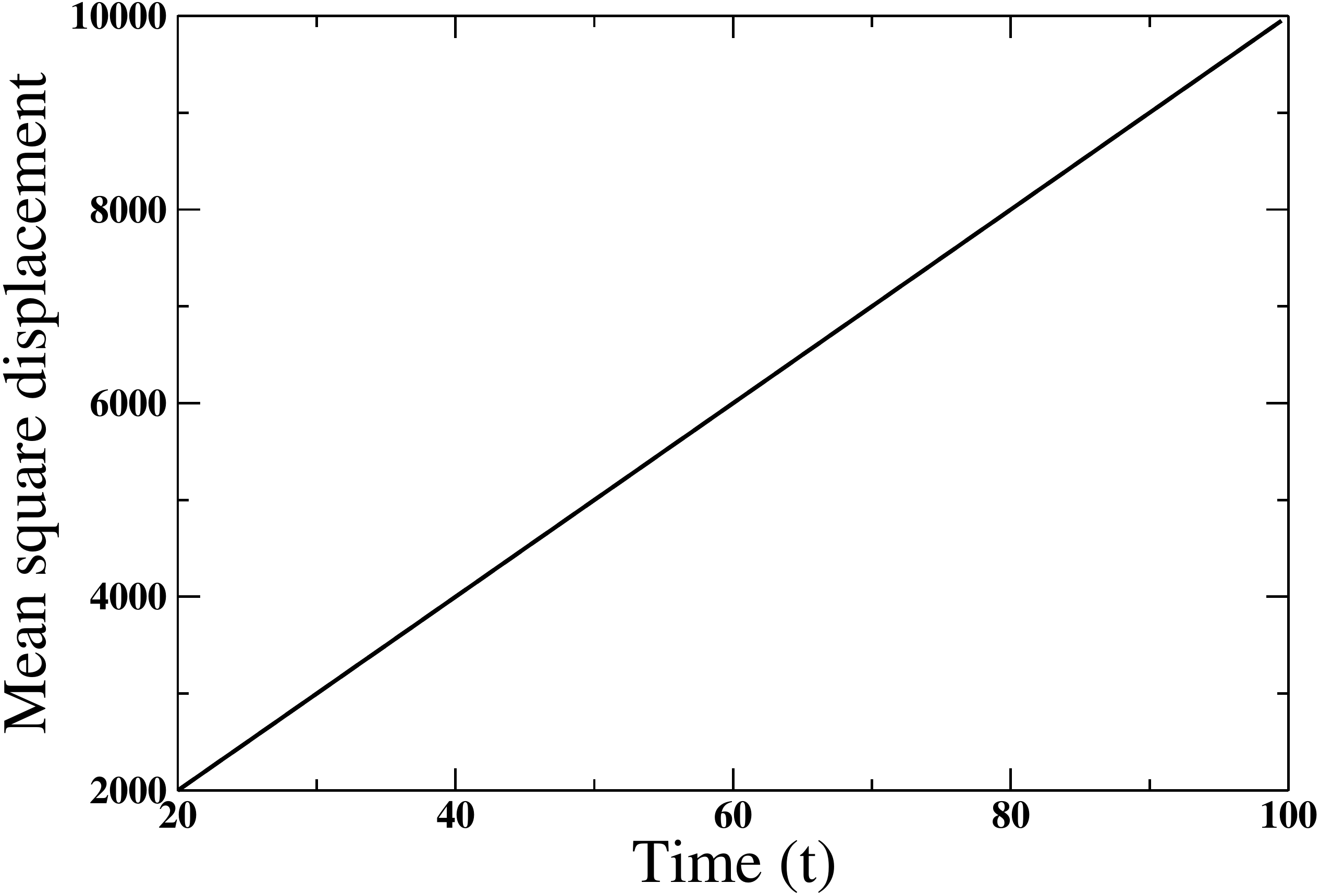}
  \caption{\label{fig7}
  Plot of the mean square displacement as a function of time in arbitrary units, under the
  condition, $t_{th}<<\tau<<t$. The curve is obtained by using Eq. (\ref{eq5}). In this case, the
  relaxation time is taken to be $ \tau=1 $ and the thermal time is taken to be $ t_{th}=0.02 $.}
\end{figure}
In Fig. \ref{fig7}, we have shown the plot of the mean square
displacement in the classical domain. Both the newer response
function and the step-function response function yield the same
curve.

In this case, considering the same relaxation time, i.e. $ \tau=10 \mu s$,
the observation time can be estimated to be of the order of a few
$ ms $ to a few $ s $ and the temperature of the order of a few $ \mu K $ to $ mK $.

One sees in the figures
that the mean square displacement
$ {\langle\Delta x^{2}\rangle} $
is positive in all six cases.

\section{Positivity conditions}
In Ref. \cite{rafsup} we had noticed that the expression for the
mean square displacement gets to be self-contradictory in a time-regime
$t\approx\tau$.  This stemmed from the fact that the response function
contemplated there did not satisfy certain positivity requirements
which we elaborate in this section.

The first such requirement is Wightman positivity,
which one could think of as a strengthened form of positivity
of the mean square displacement.
The (two-point) Wightman function $ W(t) $ is defined as,
\begin{eqnarray}
    W(t) = \left\langle x(t) \, x(0) \right\rangle
\end{eqnarray}
Wightman positivity requires it to be of positive type (also called
``positive definite''), which is equivalent to positivity of the Fourier
transform: $ \widetilde{W}(\nu)>0$.

Using the alternative (KMS-like) form of the FDT,
\begin{eqnarray}
  \widetilde{{W}}(-\nu) = e^{2\pi\beta\hbar\nu} \; \widetilde{{W}}(\nu) \ ,
\label{basic-fdr}
\end{eqnarray}
one can write for the Wightman function in the frequency domain,
\begin{eqnarray}
   \widetilde{{W}}(\nu)=\frac{1}{1-e^{2\pi\beta\hbar\nu}}(\widetilde{{W}}(\nu)-\widetilde{W}(-\nu))\label{neq6}
\end{eqnarray}
The response function defined in Eq. (\ref{e11})
can, for $ t>0 $,
be expressed in terms of the Wightman function:
\begin{eqnarray}
  R(t) = \frac{i}{\hbar} (W(t) - W(-t)) \nonumber\\
  \text{or,  }\:  W(t) - W(-t) = -i\hbar R(t) \nonumber
\end{eqnarray}
Therefore
we have for all $t$, and
for the equivalent odd function $ \Rch (t) $ of Eq. (\ref{e13})
\begin{eqnarray}
  W(t)-W(-t)&=&-i\hbar \Rch (t)\nonumber
\end{eqnarray}
or in the frequency domain,
\begin{eqnarray}
   \widetilde{W}(\nu)-\widetilde{W}(-\nu)&=&i\hbar\widetilde{\Rch }(\nu)\label{neq7}
\end{eqnarray}
Using Eqs. (\ref{neq6})  and (\ref{neq7}),
one can express $\widetilde{W}(\nu)$ in terms
of $\widetilde{\Rch}(\nu)$ as follows:
\begin{eqnarray}
  \widetilde{W}(\nu) = \frac{i\hbar} {1-e^{2\pi\beta\hbar\nu}} \ \widetilde{\Rch }(\nu) \label{n1}
\end{eqnarray}
The Wightman function $ W(t) $ is therefore of positive type if and only if
the R.H.S of (\ref{n1}) is positive for every $ \nu $.

Let us verify Wightman positivity for the response function of Eq.~(\ref{eq2}),
$R(t) =\mu \left( 1-e^{-\frac{t}{\tau}}\right)\theta(t)$.
We have then, for $ t>0 $,
\begin{eqnarray}
   \Rch (t)&= &\mu \left[ \mathrm{sgn}(t)\left( 1-e^{-\frac{\vert t\vert}{\tau}}\right)\right] \nonumber\\
   &=& \mu \left[\mathrm{sgn}(t)-\theta(t)e^{-\frac{t}{\tau}}+\theta(-t)e^{\frac{t}{\tau}}\right]
\end{eqnarray}
Therefore,
\begin{eqnarray*}
i\widetilde{\Rch }(\nu)&=&\mu \left[ i\int_{-\infty}^{\infty} dt\, e^{2\pi i\nu t}\mathrm{sgn}(t)
                                     -i \int_{-\infty}^{\infty} dt\, e^{2\pi i\nu t}\theta(t) e^{-\frac{t}{\tau}} \right. \nonumber\\
                              &+&\left. i\int_{-\infty}^{\infty} dt\, e^{2\pi i\nu t}\theta(-t)e^{\frac{t}{\tau}}\right] \nonumber\\
  &=&\mu \left[ i\int_{0}^{\infty} dt\, (e^{2\pi i\nu t}-e^{-2\pi i\nu t})
               -i\int_{0}^{\infty} dt\, (e^{2\pi i\nu t}\right.\nonumber\\
  &-&\left. e^{-2\pi i\nu t})e^{-\frac{t}{\tau}}\right]\nonumber\\
  &=&\mu \left[ -2 \mathrm{Im}\int_{0}^{\infty} dt\, e^{2\pi i\nu t} + 2 \mathrm{Im}\int_{0}^{\infty} dt\, e^{(2\pi i\nu t)-\frac{t}{\tau}}\right]\nonumber\\
  &=&\mu \left[ -2 \mathrm{Im}\left( \frac{1}{2}\delta(\nu)+\frac{i}{2\pi\nu}\right) +2 \mathrm{Im}
  \frac{1}{\left( \frac{1}{\tau}-\left( 2\pi i\nu \right) \right) }\right]\nonumber\\
  &=&\mu\left[ \frac{-2}{2\pi\nu}+2 \mathrm{Im} \frac{\frac{1}{\tau}+2\pi i\nu}{\frac{1}{\tau^{2}}+(2\pi\nu)^{2}}\right]\nonumber\\
  &=&\frac{4\pi\nu\mu}{\frac{1}{\tau^{2}}+(2\pi\nu)^{2}}-\frac{2\mu}{2\pi\nu}\nonumber\\
  &=&\frac{-\mu\frac{1}{\tau^2}}{\pi\nu\left( \frac{1}{\tau^{2}}+(2\pi\nu)^{2}\right) }
\end{eqnarray*}
Hence,
\begin{eqnarray}
\widetilde{W}(\nu)&=&\left( \frac{\hbar}{1-e^{2\pi\beta\hbar\nu}}\right) \frac{-\mu\frac{1}{\tau^{2}}}{\pi\nu\left( \frac{1}{\tau^{2}}+(2\pi\nu)^{2}\right)}\nonumber\\
&=&\frac {\hbar\mu} {\pi\nu\left(e^{2\pi\beta\hbar\nu}-1\right) \left(1 + (2\pi\nu\tau)^{2}\right) }\\
&\geq & 0
\end{eqnarray}
Therefore this response function satisfies Wightman positivity.

The second positivity requirement is passivity,
which, at linear order,
can be stated as follows \cite{ford, thirring}.
The mean work done on the system is given at this order by,
\begin{eqnarray}
  \overline{W}&=&\int dt \  f(t) \ \langle\dot{x}(t)\rangle \label{eq8}
\end{eqnarray}
where $ f(t) $ is a weak perturbing force applied to the displacement $x(t)$.
Passivity is then the requirement,
\begin{eqnarray}
  \overline{W}&\geq &0 \label{eq12}
\end{eqnarray}
(We have used the notation $ \overline{W} $ for work to distinguish it
 from the Wightman function $W$.)

By definition of the response function,
we have
\begin{eqnarray}
  \langle x(t)\rangle -\langle x(0)\rangle&=&\int dt' R(t-t')f(t')\label{eq9}
\end{eqnarray}
Hence the expression for work reduces to,
\begin{eqnarray}
\overline{W}&=&\int_{-\infty}^{\infty}dt f(t)\frac{d}{dt}\left\lbrace \int_{-\infty}^{\infty}dt'R(t-t')f(t')\right\rbrace\\
  &=& \int_{-\infty}^{\infty}dt \; f(t) \; K(t-t') \; f(t') \; dt' \ ,
  \label{eq10}
\end{eqnarray}
where $K(t)$ is the
time derivative of the position response function $R(t)$,
i.e. the response function for the velocity.
Taking
Fourier transforms,
and using the fact that both $\widetilde{K}$ and $\widetilde{f}$ are Fourier transforms of real functions,
we can write,
\begin{eqnarray*}
  \overline{W}&=& \int_{-\infty}^{\infty}dt f(t) \int_{-\infty}^{\infty}dt' f(t')    \nonumber\\
  &&\int_{-\infty}^{\infty}d\nu e^{ 2\pi i\nu\left(t-t' \right) }\widetilde{K}(\nu)^* \nonumber\\
  &=&\int_{-\infty}^{\infty}d\nu \widetilde{K}(\nu)^* \int_{-\infty}^{\infty}dt e^{ 2\pi i\nu t} f(t) \nonumber\\
  &&\int_{-\infty}^{\infty}dt' e^{-2\pi i\nu t'} f(t')                                              \nonumber\\
  &=&\int_{-\infty}^{\infty}d\nu \widetilde{K}(\nu) \vert \widetilde{f}(\nu)\vert^{2}             \nonumber\\
  &=& 2 \int_{0}^{\infty} d\nu \; \Re \; \widetilde{K}(\nu) \; \vert \widetilde{f}(\nu)\vert^{2} \\
\end{eqnarray*}
Since this must be positive for arbitrary (real) $f(t)$,
passivity at linear order reduces to the positivity of the real part of the
Fourier-transformed velocity-response function:
\begin{eqnarray}
   \Re \; \widetilde{K}(\nu) \ge 0  \label{pass}
\end{eqnarray}

One might wonder why passivity concerns only the real part of
$\widetilde{K}$ whereas Wightman positivity requires that the full
Fourier transform $\widetilde{W}$ be non-negative.  The difference is
that passivity requires positivity only for real force-functions $f$,
whereas Wightman positivity requires that $f^* W f$ be positive for
arbitrary complex functions $f(t)$.  If we treat $W(t-t')$ and $K(t-t')$
formally as matrices then, because $fKf=fK^Tf$ ($K^T$ being the
transpose), only the symmetric part of $K$ influences the work done.
Positivity of the latter then equates to positivity of the Fourier
transform of this symmetric part, which is exactly the real part of the
Fourier transform of $K$ itself.

Let us check that the requirement (\ref{pass}) is met by our response function,
$R(t) =\mu \left( 1-e^{-\frac{t}{\tau}}\right)\theta\left( t\right)$.
For this response function,
\begin{eqnarray*}
  K(t)
  &=& \frac{d}{dt}\left\lbrace \mu \left( 1-e^{-\frac{t}{\tau}}\right)\theta\left( t\right)\right\rbrace
   = \frac{\mu}{\tau} e^{-\frac{t}{\tau}} \theta(t)
\end{eqnarray*}
The Fourier transform of $ K(t) $ is then
\begin{eqnarray}
  \widetilde{K}(\nu) &=&\
  \mu \int_{-\infty}^{\infty}dt e^{2\pi i\nu t}\frac{d}{dt}\left\lbrace\left( 1-e^{-\frac{t}{\tau}}\right)\theta\left( t\right)\right\rbrace \nonumber\\
  & = & \mu \frac{\frac{1}{\tau}}{\frac{1}{\tau} - 2\pi i\nu} \ , \nonumber\\
 \label{eq13}
\end{eqnarray}
The real part of this is
\begin{eqnarray}
     \frac{ \mu  } {1 + (2\pi\nu\tau)^{2}}  \ ,
\end{eqnarray}
which is indeed non-negative for all $\nu$.

Our response function thus satisfies both positivity conditions.
This is to be contrasted with the case of the step function
response function \cite{rafsup} $R(t)=\mu\theta(t-\tau)$
where positivity fails in the limit $ t\rightarrow \tau $.

\subsection*{Positivity and the FDT}
The conditions for Wightman positivity and passivity are related by the
FDT in its different guises, (\ref{e1}) and (\ref{basic-fdr}).  By
combining these with the equation, $K=dR/dt$, one can relate $W(\nu)$ to
the real part of $K(\nu)$, as follows.

Let us begin with $\Re\Ktilde(\nu)$.
Because differentiation in the time-domain corresponds to multiplication by
$\nu$ in the frequency-domain,
we can trade $\Re\Ktilde(\nu)$ for $\Im\Rtilde(\nu)$.
The latter however, is equivalent
by (\ref{e1})
to $\Ctilde(\nu)$,
which in turn is by definition half of $\Wtilde(\nu)+\Wtilde(-\nu)$.
Then with the help of (\ref{basic-fdr}),
we can eliminate $\Wtilde(-\nu)$ from this sum
to be left with a simple multiple of $\Wtilde(\nu)$.
Following these steps,
one finds straightforwardly that
\begin{eqnarray}
  \Re\,\Ktilde(\nu) = \frac{\pi \nu}{\hbar} \; (\exp\{2 \pi \beta \hbar\nu\} - 1) \; \Wtilde(\nu) \ ,
\end{eqnarray}
which makes it evident that
$\Re\Ktilde(\nu)$ is positive if and only if $\Wtilde(\nu)$ is positive
(where we ignore, if need be, the special case $\beta\nu=0$).
Thus Wightman positivity implies linear-order passivity and conversely,
as a consequence of the FDT.



\smallskip
\noindent{\bf REMARK: }
The requirement of Wightman positivity is quite general.  Because it
merely reflects the positivity of the Hilbert space inner product, it
applies to any system whose description conforms to the quantum
formalism based on Hilbert space.  The requirement of passivity on the
other hand, reflects a very special property of systems in thermal
equilibrium, namely that one cannot extract work from them by purely
mechanical means.  It is therefore noteworthy that we have here derived
passivity simply from Wightman positivity and the FDT.  This indicates
that the latter manages to encapsulate a surprisingly large part of the
meaning of thermal equilibrium.

\section{Conclusion}
In this paper, proceeding solely on the basis of the
fluctuation-dissipation theorem (FDT) and a choice of functional form
for the response-function $R$, we have analysed the growth of mean
square displacement as a function of time $t$.  The response-function we
have used depends on two parameters, a ``mobility'' $\mu$ and a
``relaxation-time'' $\tau$, and correspondingly one encounters six
different regimes defined by the ordering among the numbers, $t$,
$\tau$, and the thermal-time $t_{th}={\beta\hbar}/{\pi}$.
(The mobility enters only as an overall prefactor.)

One encounters in all, three qualitatively different growth-laws,
which could be termed ``classical'', ``quantum'' and ``intermediate''.
When $t\gg{t_{th}}$, one recovers the linear growth familiar from
classical diffusion driven by thermal fluctuations.  When, on the other
hand, $t\ll t_{th}$ but $t\gg\tau$, one is in the properly quantum
regime of logarithmic growth driven by quantal fluctuations.
Intermediate between these cases is one where $t$ falls below both $t$
and $t_{th}$ and one encounters an intermediate growth proportional to
$t\ln{t}$.


In an earlier study \cite{rafsup}, the response function was chosen to
be a simple step function.  Such a function works well for times longer
than the relaxation-time, but for very short times, it leads to
inconsistencies stemming from the fact that the step function violates
certain positivity conditions that a putative response function must
satisfy, namely {\it\/Wightman positivity\/} (which trivially guarantees
positivity of the mean-square displacement) and the thermodynamic
condition of {\it\/passivity\/}.

In this connection we have exhibited some relationships among the
positivity conditions in question, most importantly that (for a weak
perturbing force) Wightman positivity also implies passivity when
combined with the FDT.  Indeed, we have shown that in the presence of
the FDT, linear-order passivity is equivalent to Wightman positivity.

The response function used in our present study has a twofold advantage.
Firstly, it has a form which is closer to one realizable in a cold-atom
laboratory. Secondly, it satisfies the physically mandated positivity
requirements and therefore gives theoretically consistent results in the
entire time domain. This has allowed us to go beyond Ref. \cite{rafsup}
in probing the short-time regime where $t<<\tau$.  On the other hand, in
what we have called the quantum regime, we find qualitatively the same
logarithmic growth as earlier, suggesting that this behavior is robust.
(This is not to say, however, that there are not quantitatively distinct
predictions.)



\textit{Experimental Prospects}:
The quantum law of diffusion predicted by our analysis can be tested in
experiments with ultra-cold atoms
\cite{Anderson, chu, Ketterle, Magnetic, Optical, Ramansideband, coldexpt}.
In recent years there has been considerable development in this area,
and one can now hope to probe the growth of mean square displacement as a function
of time in the time-temperature domains discussed here. For example
using sub-Doppler cooling in optical molasses it is possible to achieve
temperatures of the order of $ 10 $ micro Kelvin
\cite{chu} using laser cooling techniques.  With this technique, the
atoms are cooled and confined in a very small region of space thanks to
damping of atomic velocities.  Within the confined region, the atomic
motion is analogous to that of a Brownian particle.  Furthermore the
technique of evaporative cooling in conservative traps \cite{Anderson}
can reach temperatures of the order of $ 10 $ nano Kelvin.

Since the various time-temperature regimes discussed in this paper all
appear to be realizable in cold-atom laboratories, we are optimistic
that experiments
in the quantum and intermediate regimes will be performed soon, perhaps
by an experimental group with whom we have discussed our results.


\section{Acknowledgements}
It is a pleasure to thank Sanjukta Roy for discussions on the experimental
aspects of this work.

\noindent
This research was supported in part by NSERC through grant RGPIN-418709-2012.
This research was supported in part by Perimeter Institute for
Theoretical Physics. Research at Perimeter Institute is supported
by the Government of Canada through Industry Canada and by the
Province of Ontario through the Ministry of Economic Development
and Innovation.

\renewcommand{\theequation}{A-\arabic{equation}}
  \setcounter{equation}{0}  
\section{Appendix: Details of calculations} \label{A}

Substituting $ R(t') $ from Eq. (\ref{eq2}) in Eq. (\ref{eq1}), we get,
\begin{eqnarray}
 \langle \Delta x^2 \rangle &=&\frac{\mu}{\beta}\int_{0}^{\infty}dt'(1-e^{\frac{t'}{\tau}})\left\lbrace 2 \coth \left(\frac{t'}{t_{th}}\right)  \right.  \nonumber\\
   &-& \left. \coth \left(\frac{t'+t}{t_{th}}\right)-\coth \left( \frac{t'-t}{t_{th}}\right) \right\rbrace\nonumber\\
  &=& \lim_{\epsilon\rightarrow 0} (I_1 -I_2)
  \label{a1}
\end{eqnarray}
\begin{eqnarray}
  I_1&=&
    \frac{\mu}{\beta}\int_{\epsilon}^{\infty}dt'\left\lbrace 2 \coth \left(\frac{t'}{t_{th}}\right) - \coth \left(\frac{t'+t}{t_{th}}\right)\right. \nonumber\\
  &-&\left. \coth \left( \frac{t'-t}{t_{th}}\right) \right\rbrace \label{a2}\nonumber\\
  &=&  \lim_{t_{\infty}\rightarrow \infty}\frac{\mu }{\beta}t_{th}\left\lbrace
    2 \ln\left[ \sinh\left(\frac{t'}{t_{th}}\right)\right] -\ln \left[ \sinh\left(\frac{\vert t'+t\vert}{t_{th}}\right)\right]  \right. \nonumber\\
  &-&\left. \ln \left[ \sinh\left( \frac{\vert t'-t\vert}{t_{th}}\right)\right] \right\rbrace_{\epsilon}^{t_{\infty}} \label{a3}\nonumber\\
  &=&  \frac{2\mu}{\beta}t_{th}\ln
   \left[ \frac{\sqrt{\sinh\left( \frac{\vert t+\epsilon\vert}{t_{th}}\right)\sinh\left( \frac{\vert t-\epsilon\vert}{t_{th}}\right)  }}{\sinh\left( \frac{\epsilon}{t_{th}}\right) }\right]\nonumber\\
  &=&  \frac{2\mu}{\beta}t_{th}\ln \left[ \frac{\sinh\left(\frac{ t}{t_{th}}\right)}{\sinh\left( \frac{\epsilon}{t_{th}}\right) }\right]\label{a4}
\end{eqnarray}
using, $ t<<t_{\infty} $.
\begin{eqnarray}
I_2&=&  \frac{\mu}{\beta}
  \int_{\epsilon}^{\infty}dt'e^{-\frac{t'}{\tau}}\left\lbrace 2 \coth \left(\frac{t'}{t_{th}}\right) - \coth \left(\frac{t'+t}{t_{th}}\right)\right.  \nonumber\\
&-&\left. \coth \left(\frac{t'-t}{t_{th}}\right) \right\rbrace \label{a5}
\end{eqnarray}
\begin{eqnarray}
\text{Consider,}\: I&=&\int_{\epsilon}^{\infty}dt'e^{-\frac{t'}{\tau}}\coth \left(\frac{t'}{t_{th}}\right)\label{a6}
\end{eqnarray}
Setting $ \coth\left(\frac{t'}{t_{th}}\right)=y $, then the above integral reduces to,
\begin{eqnarray}
I&=&-t_{th}\int_{y_0}^{1}dy \left( \frac{y-1}{y+1}\right)^{\frac{t_{th}}{2\tau}-1} \frac{y}{(y+1)^2}\label{a7}
\end{eqnarray}
where, $ y_0=\coth\left(\frac{\epsilon}{t_{th}}\right) $. Now, if we substitute, $ (y-1)/(y+1)=zx $, such that, $ z=(y_0-1)/(y_0+1)=e^{-\frac{2\epsilon}{t_{th}}} $, then this integral reduces to,
\begin{eqnarray}
I&=&-z^\frac{t_{th}}{2\tau}t_{th}\int_{0}^{1}dx  x^{\frac{t_{th}}{2\tau}-1}\left( \frac{1}{2}-\frac{1}{1-zx}\right)\label{a8}
\end{eqnarray}
Using the integral form of Hypergeometric Function $ {}_2 F_1 $,
\begin{eqnarray}
{}_{2}F_{1}(a,b,c,z)&=&\frac{\Gamma(c)}{\Gamma(b)\Gamma(c-b)}\int_{0}^{1}dx x^{b-1}(1-x)^{c-b-1}\nonumber\\
&&(1-zx)^{-a}\label{a9}
\end{eqnarray}
we can write Eq. (\ref{a8}) as,
\begin{eqnarray}
I&=&-\tau e^{-\frac{\epsilon}{\tau}}\nonumber\\
&&\left[ 1-2{}_{2}F_{1}\left(1, \frac{t_{th}}{2\tau},1+\frac{t_{th}}{2\tau},e^{-\frac{2\epsilon}{t_{th}}}\right) \right] \label{a10}
\end{eqnarray}
We use the following forms of the Hypergeometric functions in our analytical calculations:
\begin{eqnarray}
{}_{2}F_{1}\left(a,b,a,z\right)&=&(1-z)^{-b}    \label{a10a}\\
{}_{2}F_{1}\left(a,b,a,0\right)&=&1 \label{a10b}\\
{}_{2}F_{1}\left(a,b,b,z\right)&=&(1-z)^{-a} \label{a10c}\\
{}_{2}F_{1}\left(a,b,b,0\right)&=&1     \label{a10d}
\end{eqnarray}
Similarly, we can evaluate the other integrals in Eq. (\ref{a5}) and get,
\begin{eqnarray}
I_2&=&  \frac{2\mu\tau}{\beta}e^{-\frac{\epsilon}{\tau}}\left[ 2{}_{2}F_{1}\left(1, \frac{t_{th}}{2\tau},1+\frac{t_{th}}{2\tau},e^{-\frac{2\epsilon}{t_{th}}} \right)\right.\nonumber\\
&-&{}_{2}F_{1}\left(1, \frac{t_{th}}{2\tau},1+\frac{t_{th}}{2\tau},e^{-\frac{2(t+\epsilon)}{t_{th}}} \right)\nonumber\\
&-&\left.{}_{2}F_{1}\left(1, \frac{t_{th}}{2\tau},1+\frac{t_{th}}{2\tau},e^{-\frac{2(t-\epsilon)}{t_{th}}} \right)\right] \nonumber\\
&=&  \frac{4\mu\tau}{\beta}\left[ {}_{2}F_{1}\left(1, \frac{t_{th}}{2\tau},1+\frac{t_{th}}{2\tau},e^{-\frac{2\epsilon}{t_{th}}} \right)\right.\nonumber\\
&-&\left.{}_{2}F_{1}\left(1, \frac{t_{th}}{2\tau},1+\frac{t_{th}}{2\tau},e^{-\frac{2 t}{t_{th}}} \right)\right]\label{a11}
\end{eqnarray}
Therefore using Eqs. (\ref{a4}) and (\ref{a11}), mean square displacement can be written as,
\begin{eqnarray}
\langle \Delta x^2 \rangle &=& \lim_{\epsilon\rightarrow 0} \frac{2\mu}{\beta }t_{th}\Biggl\lbrace \ln \left[ \frac{\sinh\left(\frac{ t}{t_{th}}\right)}{\sinh\left( \frac{\epsilon}{t_{th}}\right) }\right]\Biggl. \nonumber\\
&-&\frac{2\tau}{t_{th}}\left[ {}_{2}F_{1}\left(1, \frac{t_{th}}{2\tau},1+\frac{t_{th}}{2\tau},e^{-\frac{2\epsilon}{t_{th}}}\right)\right. \nonumber\\
&-&\left.{}_{2}F_{1}\left(1, \frac{t_{th}}{2\tau},1+\frac{t_{th}}{2\tau},e^{-\frac{2 t}{t_{th}}} \right) \right] \Biggl.\Biggl\rbrace \label{a12}
\end{eqnarray}

In this expression for mean square displacement, the logarithmic and Hypergeometric functions of
$ \epsilon $ diverge, when $ \epsilon\rightarrow 0 $.
These divergences cancel if we expand the Hypergeometric Function for small $ \epsilon $:
\begin{eqnarray}
&&{}_{2}F_{1}\left(1, \frac{t_{th}}{2\tau},1+\frac{t_{th}}{2\tau},e^{-\frac{2\epsilon}{t_{th}}} \right)\nonumber\\
&\approx &   -\frac{t_{th}}{2\tau}\left( \psi ^{(0)}\left(\frac{t_{th}}{2\tau}\right)+\ln\left(\frac{2\epsilon}{t_{th}} \right) +\gamma\right) \nonumber\\
&-&\frac{\epsilon t_{th}}{2\tau ^2}\left(-\frac{\tau}{t_{th}} +\psi^{0}\left(1+ \frac{t_{th}}{2\tau}\right)+\ln\left(\frac{2\epsilon}{t_{th}} \right)+\gamma -1\right)\nonumber\\
&-&\frac{\epsilon ^2 t_{th}}{24\tau ^3  }\left(\frac{2\tau ^2}{t_{th}^2} -6 \frac{\tau}{t_{th}} +6 \psi^{0}\left(1+ \frac{t_{th}}{2\tau}\right)\right. \nonumber\\
&+& \left. 6\ln\left(\frac{2\epsilon}{t_{th}} \right)+6 \gamma -9 \vphantom{\frac{1}{2}}\right)+ O\left(\epsilon ^3\right)+.... \label{a13}
\end{eqnarray}
where, $ \psi^{0}(x) $ is Polygamma function of order zero and $ \gamma $ is Euler-Mascheroni constant.
The last step used the relations,
\begin{eqnarray}
\psi^{0}(x)=x\sum_{n=1}^{\infty}\frac{1}{n(n+x)}-\frac{1}{x}-\gamma
\end{eqnarray}

Substituting Eq.~(\ref{a13}) in Eq.~(\ref{a12}), we get finally,
\begin{eqnarray}
\langle \Delta x^2 \rangle &=&\frac{2\mu}{\beta}t_{th}\left\lbrace \ln\left[2\sinh\left(\frac{t}{t_{th}}\right)\right]
+\psi^{0}\left(1+ \frac{t_{th}}{2\tau}\right)+\gamma +\right.\nonumber\\
&& \left.\frac{2\tau}{t_{th}} \left[{}_{2}F_{1}\left(1, \frac{t_{th}}{2\tau},1+\frac{t_{th}}{2\tau},e^{-\frac{2 t}{t_{th}}}\right)\right.\right.\nonumber\\
&-& \left. \left. 1\vphantom{\frac{1}{2}} \right] \right\rbrace \label{a14}
\end{eqnarray}
using the identity:
\begin{eqnarray}
\psi^{0}\left(1+x\right)=\psi^{0}\left(x\right)+\frac{1}{x}
\end{eqnarray}
The asymptotic forms of $ \psi^{0}\left(1+x\right) $ are:
\begin{eqnarray}
\psi^{0}\left(1+x\right)&=&\ln(x),\; \;x>>1 \label{a15}
\end{eqnarray}
\begin{eqnarray}
\psi^{0}\left(1+x\right)&=&-\gamma + \frac{\pi^2}{6}x,\; \; x<<1 \label{a16}
\end{eqnarray}
We have used these asymptotic limits in our analytical calculations.

\bibliography{supurnareferences}
\end{document}